# Spatially resolved ultrafast precessional magnetization reversal

W. K. Hiebert, L. Lagae,[a)] J. De Boeck

IMEC, Kapeldreef 75, B-3001 Leuven, Belgium

PACS numbers: 75.60.Jk, 78.47.+p, 76.50.+g, 85.70.Kh

Spatially resolved measurements of quasi-ballistic precessional magnetic switching in a microstructure are presented. Crossing current wires allow detailed study of the precessional switching induced by coincident longitudinal and transverse magnetic field pulses. Though the response is initially spatially uniform, dephasing occurs leading to nonuniformity and transient demagnetization. This nonuniformity comes in spite of a novel method for suppression of end domains in remanence. The results have implications for the reliability of ballistic precessional switching in magnetic devices.

---

[a] Also at ESAT, K.U.Leuven, Leuven, Belgium

Precessional switching was considered as far back as 1960 in magnetism's first foray into the random access memory market[1,2]. Intense interest has resurfaced in the current run for MRAM as the ultimate speed limit to magnetic switching has been approached[2,3,4,5,6,7,8,9,10]. In conventional switching, a magnetic field is applied in the opposite direction to the magnetization (making this direction the lowest energy state) and reversal proceeds through thermal attempt processes, domain nucleation, and wall motion. Bypassing the thermal and damping regimes, precessional switching relies solely on nonequilibrium dynamics as a fast magnetic field pulse applied perpendicularly to the magnetization induces a large angle precessional motion. By properly tailoring the length of the transverse field pulse, the motion can be terminated after $180^\circ$ of rotation, thus, effecting a magnetic switch via a large angle half precession. By fine enough tailoring, residual precession may even be suppressed[5,6] in the so-called ballistic precessional switch[9].

Precessional switching in microstructures has recently been directly observed by several groups[11,12,13,14]. However, these reports only showed response that was averaged over a portion of a microstructure or taken at spot locations. In order to properly probe the fundamental limits of this switching mode, spatially resolved measurement over the entire microstructure is a necessity. In particular, the uniformity (or lack thereof) of response will be a key issue in determining applicability for magnetic memory and sensor devices. Further, precessional switching induced by coincident orthogonal pulses, a mainstay in bit selection schemes for MRAM, has yet to be experimentally explored.

In this Letter, we report fully spatially resolved observation of precessional switching. A novel technique is used to suppress end domains in the remanent state of a 20x7 μm$^2$ microstructure and vector-resolved quasi-ballistic switching from remanence is demonstrated. The full spatial resolution shows complete uniformity of response for the initial motions, but reveals deviations developing during the precession. For complex pulsing parameters, the spatial variation becomes more pronounced until a would-be demonstration of precessional switching induced by coincident orthogonal pulses is ultimately defeated by transient demagnetization.

The experimental data is obtained using time resolved scanning magneto-optical Kerr microscopy[10,11]. Spatial resolution is diffraction limited at ~0.7μm; temporal resolution is jitter limited to nominally 50 ps by the delay generator. An optical micrograph of the device is shown in Fig. 1(a). The magnetic element sits on top (for optical access) of two crossing current carrying wires. Current pulses through the wires allow coincident transverse and longitudinal transient magnetic fields to be applied. The axes definitions given remain throughout and the initial magnetization state always points along the positive x-axis. The transverse pulse H$_y$(t) is directed to the right using a nominally 60ps risetime pulser. For solo transverse pulses, a reset pulse of ~+6.76 kA/m, 15ns long is applied to set the initial state. For combination pulses, the reset is between 1.99 and 2.39 kA/m and is sustained for about 1 $\mu$s.

Given the importance of uniformity of response for magnetic applications, an elliptical shape (an ellipsoid theoretically giving spatially uniform demagnetizing fields) is chosen for the element. The 16 nm magnetic film is sputtered with easy axis along the long axis of the ellipse, a photoresist pattern of 20x7 μm$^2$ is defined on top, and ion milling used to define the pattern. Isolated ellipses of this size such as those shown in Fig. 1(b) exhibit end domains in the remanent state. When tested in precessional switching, such microstructures respond uniformly near the center while the edges behave more chaotically.

To overcome this problem a novel method is used to improve the uniformity of magnetization in the remanent state. Instead of etching completely through during patterning, the magnetic film is etched partially leaving a mesa embedded in an infinite magnetic layer (in this case leaving about 5nm). The completely uniform remanent state of such an element is shown in Fig. 1(c). Instead of being forced to remain parallel to the sample edges, the magnetization aligns with the easy axis as flux lines of the would-be free poles flow out into the background as shown in the vector image. Though more tests need to be done, this embedding process may become relevant for devices where uniformity of a micron-size magnetic contact is crucial (and the cell of interest is now a stadium or rectangular shape where even less uniformity of isolated samples is expected).

Local spot response as a function of transverse pulse width in the center of the embedded ellipse is investigated in Fig. 2. Figure 2(a) and (b) show 3d views of the magnetization vector trajectory for two special cases of pulse width, the first (1.27 kA/m and 400 ps FWHM) resulting in a half-precession quasi-ballistic switch, the second (2.23 kA/m and 730 ps FWHM) in a full-precession "non-switch". Both trajectories begin with relaxive motion along the y-axis until the slew rate of the pulse rising edge is high enough to excite precessional motion. The torque induced by the cross product of M and H then tilts the vector up out of the x-y plane. The resulting demagnetizing field (pointing straight down) then dominates the effective field contributions causing most of the precession to be around this (z) axis. For the quasi-ballistic case, the applied field decays after the magnetization crosses the y-z plane resulting in a permanent switch. The smooth, "ballistic" trajectories of the projections inform the choice of name for this switching mode. For the other case, the longer pulse causes one full oscillation around the y-axis, $m_x$ changing sign twice, and magnetization is left in the unswitched state at termination.

These trajectories are broken down into normalized component form $m_x$, $m_y$, and $m_z$[15] as a function of time in Fig. 2(c) and (d), respectively (thin traces with symbols). Corresponding symbol-less solid traces are from a macrospin simulation using actual field pulse shapes[16]. The field pulses are dotted curves (though positive, shown below the axis to avoid cluttering the data) and the heavy trace tracks the measured magnetization vector length giving an estimate of the error in the measurement technique, as well as hinting at temporal moments where the magnetization is changing quickly (either as a function of space or time) with respect to the measurement resolution.

For the ballistic case (Fig. 2(c)), $m_x$ falls smoothly from positive saturation to negative saturation in 320 ps with simulated response closely following. As seen in the trajectory, the motion is almost perfectly terminated with little ringing of the measured components. Termination is assisted by a larger apparent damping (a signature of spin wave generation in these large precessions) and the small damping assumed in the simulation leads to overly strong oscillations in $m_y$.

The right vertical dashed line through Fig. 2(c-d) highlights the primary cause of the quasi-ballistic trajectory. Placed where the $m_z$ component crosses zero, this is the point[9] where the pulse field must shut off to properly terminate the dynamic motion and suppress most of the ringing. Although with finite shut-off time, the field pulse in Fig. 2(c) essentially meets this condition. In the non-switching case, $m_x$ dips smoothly and quickly from positive to near negative saturation before immediately returning to the initial state since the field pulse has remained on as $m_z$ crosses zero. Accounting for the differences in damping, the simulated data matches fairly well implying that the macrospin model is useful for describing behaviour in the center of the sample.

Further pulse width dependence is explored in Fig. 2(e) where an oscillation in the final state with increasing pulse width is seen. Each row is a vector plot versus time for the measured magnetization with a grayscale background of the simulated magnetization angle. Onset of the transverse pulse is indicated by the vertical dashed-line and termination approximately indicated by the white dot.

The termination of the pulse determines whether the macrospin finishes in a switched state or not (that is, whether the field is removed after an odd or even number of half oscillations). For the first two rows, the vector does not cross the y-z plane and no switching takes place. For the next three, the magnetization switches, indicated by both the white band and final direction of the arrows. Note the "ringing" in the $3^{rd}$ and $5^{th}$ rows where termination is too early and too late, respectively. By the $6^{th}$ row, the longer pulse has allowed a $2^{nd}$ crossing of the y-z plane resulting in full precession and the no-switch final state. For the $8^{th}$ through $10^{th}$ rows, there are 3 crossings of the hard plane and 3/2-oscillation switching. Again, simulation and experiment compare well with only row 8 differing in final state.

Interesting to note is that rows 11-13 fail to complete 4/2 oscillations to end unswitched in spite of the magnetization crossing the hard plane a $4^{th}$ time. This reminds us that the sample is not truly a macrospin which would have relaxed to the unswitched direction. In fact, the measured switched cases stabilize in a canted direction after pulse termination, as opposed to along the symmetry axis.

Spatially resolved measurement becomes necessary to elucidate this issue. Figure 3 shows a montage of grayscale images (250 ps intervals) of the instantaneous magnetization for a transverse pulse case of 3/2 oscillations of the center of the structure. The magnetization looks uniform[17] until frame 6, where the pulse begins to turn-off. With the Zeeman term removed, the y-component relaxes to zero and the vector turns to whichever easy-axis is closer; the left and right edges align with the positive x-axis while the center tilts toward the negative x-axis (more easily seen in the vector movie[18]). This implies there must have been a spatial variation in internal field due to nonuniform magnetostatic energy leading to a subsequent dephasing process for the precessional oscillations across the sample.

This can be understood more clearly viewing the grayscale linecut (horizontally across the sample) vs. time shown below the montage. Starting from complete uniformity, a "C" shaped curvature develops, becoming more pronounced over time, indicating that response at the edges lags behind that at the center. A similar dephasing occurs for other pulse lengths with greater spatial variation tending for longer pulses and more crossings of the hard plane. In spite of the immersion magnetic layer and the elliptical shape, which should both moderate (and homogenize) the internal magnetic field, it thus seems that magnetostatic waves still contribute and that these transverse precession-inducing pulses can "transiently demagnetize" the sample.

This dephasing already occurs with a single transverse field pulse. However, most bit selection schemes in MRAM require overlapping orthogonal pulses to address individual cells.  Independently, each pulse is smaller than the switching threshold so that cells all along the word line and bit line are unaffected; combined, they switch the cell at the intersection of the two lines.  To see if dephasing is exacerbated, the ellipse response to coincident crossing "half-select" pulses for this precessional mode switching is shown in Fig. 4.  $H_x(t)$ is 15 ns long with a rise time of 2 ns and strength of -500 A/m.  $H_y(t)$ is roughly Gaussian in shape, of peak amplitude +716 A/m, with FWHM 360 ps.  Neither pulse causes a switch on its own; combined they should cause switching according to the Stoner-Wohlfarth astroid.  The y-pulse is applied about 4 ns after the beginning of the x-pulse.

In $m_x$, although the center achieves about 75% reversal, the edge regions remain mainly unswitched through time and make up a substantial fraction of the cell area.  More complicated yet, the $m_y$ response is split down the center of the element, the contrast implying clockwise rotation on the left half of the sample and counterclockwise rotation on the right; the magnetization begins precessing the wrong way in a sizable portion of the element!  This perplexing response cannot be caused by the out-of-plane gradient of the pulse fields (related to the finite-size of the transmission line) since the symmetry breaking direction is opposite to what one would expect from this effect.  It must be an energy-lowering demagnetizing maneuver by the element responding to some complex internal field.  The final, transiently demagnetized, metastable state is shown in Fig. 4(c) and is similar to that of Fig. 3 with a crucial difference.  The center of the sample has crossed the hard plane and is technically switched.  However, it is only held metastably in place by the -500 A/m longitudinal field; when this field is removed at the end of 15 ns, the center relaxes back (within 5 or 10 ns) to the initial unswitched direction.  The large edge regions pull the center back and the permanent switch cannot be maintained. The transient demagnetization of the sample ultimately causes precessional mode coincident-pulse writing to fail.

Spatially resolved precessional magnetic switching data have shown spatial incoherence during reversal. Dephasing of oscillations give evidence of nonuniform internal fields despite the homogenizing effects of both sample shape and embedding layer. The dephasing can lead to clear nonuniformity; with coincident magnetic field pulses, highly varying internal fields can cause complex magnetization response and metastable states. The consequence of such internal field deviation, transient demagnetization of the element, raises concern for use of precessional switching in sensor and MRAM applications.

The authors thank J. Das, R. Wirix-Speetjens, J. Bekaert, and P. Van Dorpe for technical assistance. W.K.H. acknowledges the financial support of the Natural Sciences and Engineering Research Council, NSERC (Canada). L.L. acknowledges the financial support of IWT, Flanders (Belgium).

**FIG. 1 Experimental geometry.** (a) Optical micrograph of embedded elliptical element on top of crossing current wires with pulse field directions and axes definitions. (b) Grayscale images of $m_x$ and $m_y$ for a 12x4 $\mu m^2$ isolated elliptical element showing end domains in remanence. (c) Grayscale and vector map images for the 20x7 $\mu m^2$ embedded element in remanence. The closure domain process occurs in the background magnetic material leaving the elliptical mesa highly uniformly magnetized.

**FIG. 2 Measured 3D trajectory and projections of the magnetization vector for (a) half-precession quasi-ballistic reversal and (b) full-precession non-reversal.** The solid line is the vector tip trajectory and dotted lines are its projection onto the planes. (c,d) $m_x$, $m_y$, and $m_z$ as a function of time for the data from (a) and (b), respectively. Curves with symbols are experimental data, medium weight curves are corresponding macrospin simulation data, and the bold curve is the measured magnetization vector length. The shape of the magnetic field pulse $H_y(t)$ inducing the motion is indicated by a dotted line. The right vertical dashed line indicates the zero crossing of $m_z$. (e) Pulse width dependence of the precessional reversal. Vectors track the in-plane direction of measured magnetization over time and the grayscale background gives the in-plane angle of simulated magnetization. Field pulses begin at the vertical dashed line and approximately end at the white dots. Increasing pulse-width results first in reversal, then non-reversal, then reversal again, as the hard plane is crossed once, twice, and three times, respectively.

11**FIG. 3** Montage of spatial images (250 ps spacing) of the instantaneous magnetization during a precession that crosses the hard plane three times (also in movie form[18]). Frames 6-8 of $m_x$ show clear nonuniformity corresponding with decay of the field pulse (shown just above). Below is a horizontal linecut of $m_x$ displayed in grayscale as a function of time. A curvature (at 0.7 ns) indicates phase lag of the precession at the sample edges.

**FIG. 4** (a) Montage of spatial images (200 ps spacing) of the magnetization response to coincident orthogonal magnetic field pulses. (b) Horizontal linecut vs. time; $m_y$ exhibits complex variation. (c) Metastable state before slow field pulse shut-off. Magnetization eventually returns to positive saturation.

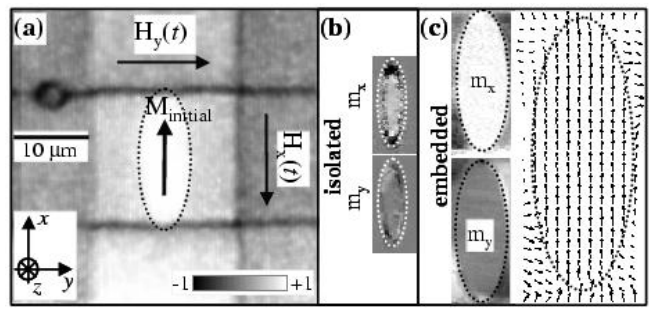

Figure 1

W. K. Hiebert *et al.*, submitted

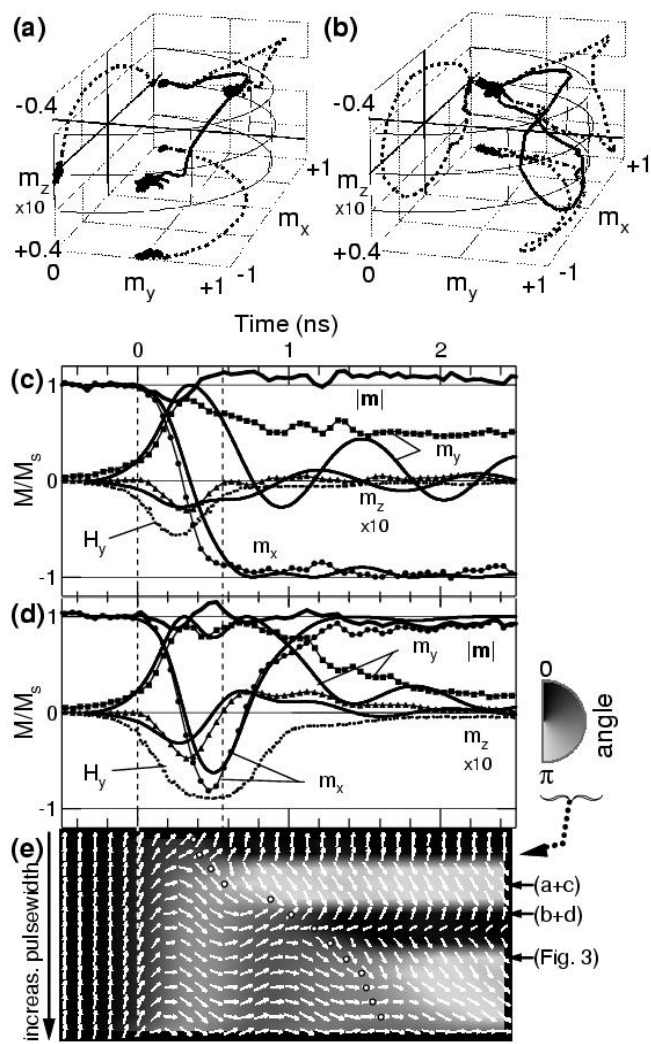

Figure 2

W. K. Hiebert *et al.*, submitted

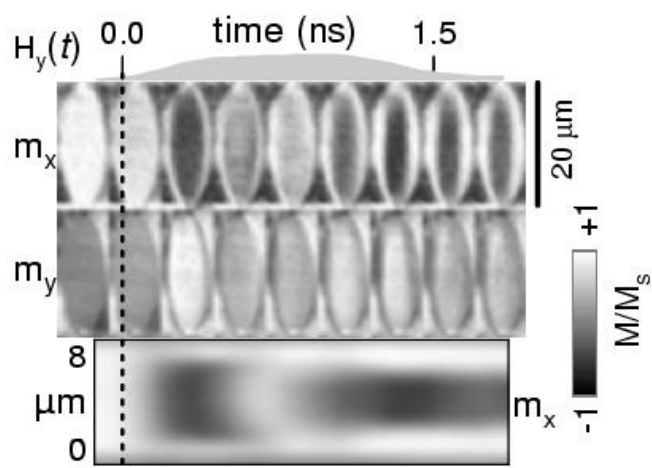

Figure 3

W. K. Hiebert *et al.*, submitted14

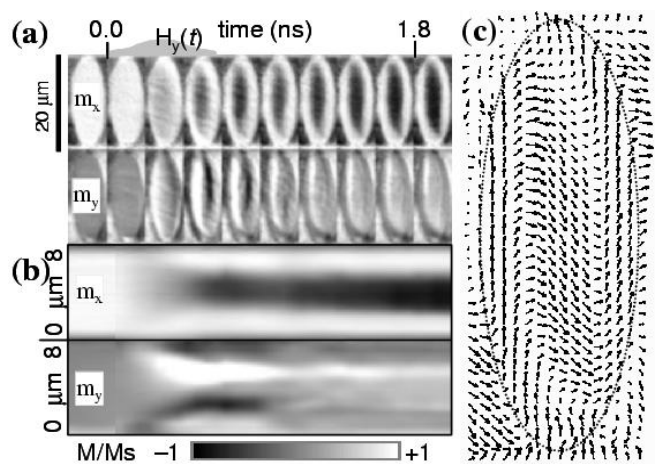

Figure 4

W. K. Hiebert *et al.*, submitted

---

[16] Macrospin parameters like Nx, Ny, Nz, etc. Hysteresis properties of the bulk (16nm) film show a coercivity of about 80 A/m with $H_k$~400 A/m.

[17] The thin band of nonuniform response around the perimeter is likely due to a loss of signal as the laser spot encounters the sample edge.

[18] Vector movie of the data to be made available on EPAPS.